\documentclass[
notitlepage,
aps,
nobibnotes,
nofootinbib]{revtex4-1}

\usepackage{amsmath}    
\usepackage{mathtools}
\usepackage{bm}         
\usepackage{graphicx}   
\usepackage{grffile}
\usepackage{fancyvrb}   
\usepackage{color}      
\usepackage{subfigure}  
\usepackage{hyperref}   
\allowdisplaybreaks[1]
\usepackage[margin=0.7in]{geometry}
\def\d{\mathrm{d}}
\def\pd{\partial}
\def\b#1{\mathbf{#1}}

\usepackage{wrapfig}
\usepackage{titlesec}
\usepackage{amssymb}

\begin{document}
\author{Matt Majic} \email{mattmajic@gmail.com}
\author{Eric C. Le Ru} \email{eric.leru@vuw.ac.nz}
\affiliation{The MacDiarmid Institute for Advanced Materials and Nanotechnology,
School of Chemical and Physical Sciences, Victoria University of Wellington,
PO Box 600, Wellington 6140, New Zealand}

\title{Relationships between solid spherical and toroidal harmonics}

\begin{abstract}
Relationships are derived expressing solid spherical harmonics as series of toroidal harmonics and vice versa. The expansions include regular and irregular spherical harmonics, ring and axial toroidal harmonics of even and odd parity about the plane of the torus. The expansion coefficients are given in terms of a recurrence relation. The existence/in-existence of these expansions is discussed in terms of the toroidal geometry and the singularities of the harmonics. The expansions are used to express the potential of a charged conducting torus on a basis of spherical harmonics. 
\end{abstract}
\maketitle
\section{Introduction}
Note: the main results of this work are also included in the appendix of \cite{majic2019toroidalTmat}, where toroidal harmonics are used to find analytical expressions for the quasistatic limit of the scattering T-matrix for a torus.

Solid\footnote{"Solid" here means the full solution to Laplace's equation. For example, solid spherical harmonics include the radial part. We will generally omit this term and it should be assumed that all harmonic functions mentioned here are solid.} spherical harmonics are a well known tool for expanding potentials in electrostatics, magnetostatics, classical gravity,  and other physical phenomia satsfying Laplaces's equation. Toroidal harmonics, which correspond to the potentials of thin ring sources, have had recent interest in application to gravitational fields of ring like astronomical structures \cite{Fukushima2016}. The gravitational potential of a solid torus has been expressed in terms of both spherical \cite{Kondratyev2012} and toroidal harmonics \cite{wong1973toroidal,majic2020surface}. Toroidal harmonics have also seen recent interest in electromagnetism, for example the magnetic field around a superconducting torus \cite{ivaska1999}, the field of a magnetized torus \cite{beleggia2009}, and low frequency acoustic or electromagnetic scattering of a point charge or dipole near a torus \cite{Venkov2007,vafeas2016torusdipole}.

Some simple expansions between toroidal and spherical harmonics are well known. Toroidal harmonics of degree zero, corresponding to the potential of rings of sinusoidal charge distributions, are known as series of spherical harmonics. Spherical harmonics corresponding to point charges and dipoles are known as series of toroidal harmonics. Less well known, relationships for all degrees and orders have been derived in a 1983 Russian paper  \cite{erofeenko1983addition}, along with relationships between toroidal and cylindrical harmonics. The relationships were subsequently used to study the electrostatic interaction of a torus and a sphere \cite{shushkevich1998electrostatic}. In this document we re-derive these expansions and express the expansion coefficients in a simpler form.


Toroidal and bispherical coordinates are closely related; they essentially differ by a real/imaginary interchange of the focal distance, and the corresponding harmonics differ by a shift of the separation constant by $\pm1/2$. Formulae relating spherical/bispherical harmonics are similar and derived in \cite{majic2019bispherical}.

The paper is organized as follows. Section 2 defines toroidal coordinates and harmonics and investigates their singularities, and presents relevant expansions of Green's function. Section 3 derives the expansions of toroidal harmonics in terms of spherical harmonics, followed by the expansions of spherical harmonics in terms of toroidal harmonics. Section 3 C discusses how the geometry of the singularities of the toroidal harmonics affects the existence of these expansions. Section 4 applies these expansions to the potential of a charged conducting torus. The appendix provides series representations of the toroidal functions. Codes for their computation are attached as supplementary material.

\section{Preliminaries}
\subsection{Toroidal coordinates}
First define spherical and cylindrical coordinates\footnote{atan2 is a similar to the arctangent but provides correct results in all four quadrants of $x$ and $y$.}:
\begin{align}
r=\sqrt{x^2+y^2+z^2}, \qquad \rho=\sqrt{x^2+y^2}, \qquad u=\cos\theta=\frac{z}{r}, \qquad \phi=\text{atan2}(y,x) 
\end{align}

Then toroidal coordinates $(\xi,\eta,\phi)$ with a focal ring radius $a$ are defined as 
\begin{align}
\xi=\frac{1}{2}\log\frac{(\rho+a)^2+z^2}{(\rho-a)^2+z^2}, \qquad 
\eta=\text{sign}(z)\text{acos}\frac{r^2-a^2}{\sqrt{(r^2+a^2)^2-4\rho^2a^2}},
\end{align} 
with ranges $\xi\in[0,\infty) ,~\eta\in(-\pi,\pi]$. $\xi$=constant defines a torus, and $\eta,\phi$ define a point on the torus surface. 
For convenience we also define:
\begin{align}
\beta&=\cosh\xi=\frac{r^2+a^2}{\sqrt{(r^2+a^2)^2-4\rho^2a^2}}=\frac{\chi}{\sqrt{\chi^2-1}} \\
\chi&=\coth\xi=\frac{r^2+a^2}{2\rho a}=\frac{\beta}{\sqrt{\beta^2-1}}, 
\end{align}
with ranges $\beta\in[1,\infty),~\chi\in[1,\infty)$. The focal ring lies at $\xi=\beta=\infty$, $\chi=1$, and both the $z$-axis and $r\rightarrow\infty$ lie at $\xi=0$, $\beta=1$, $\chi=\infty$.
On the $xy$-plane, $\sigma=0$ for $r>a$ and $\sigma=\pi$ for $r<a$. The sphere $r=a$ corresponds to $\sigma=\pm\pi$.

\begin{figure}
\includegraphics[scale=.45]{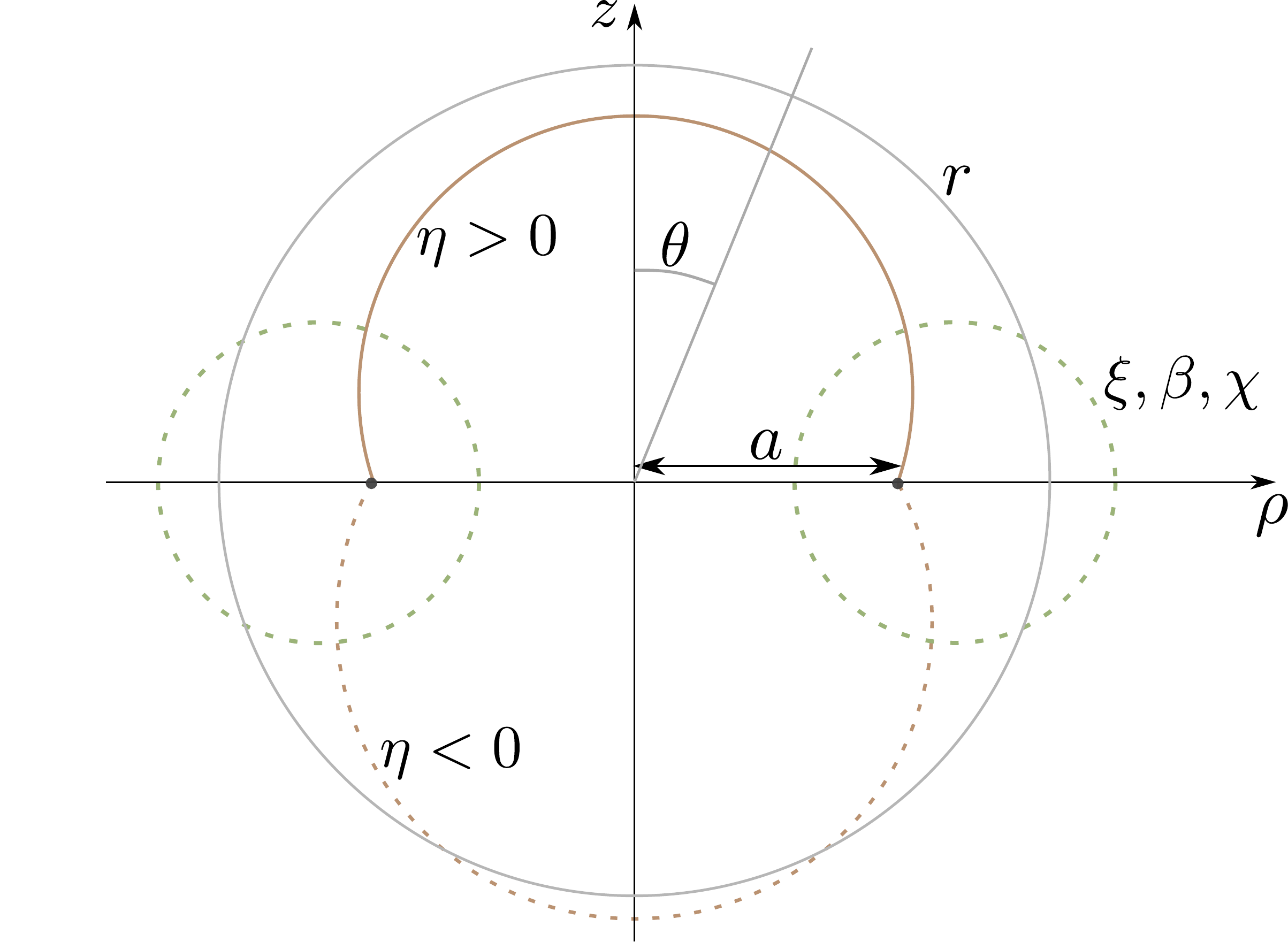}
\caption{Schematic illustrating cross sections of surfaces constant values of the relevant coordinates. $\xi,\beta,\chi$ all define a torus surface, and $\eta$ relates to the angle wrapping through the focal ring. The sphere at $r$ defines the boundary of convergence of the expansions of toroidal harmonics on a basis of spherical harmonics.}
\end{figure}

\subsection{Toroidal harmonics}

\begin{figure}[h]
	\includegraphics[scale=.63]{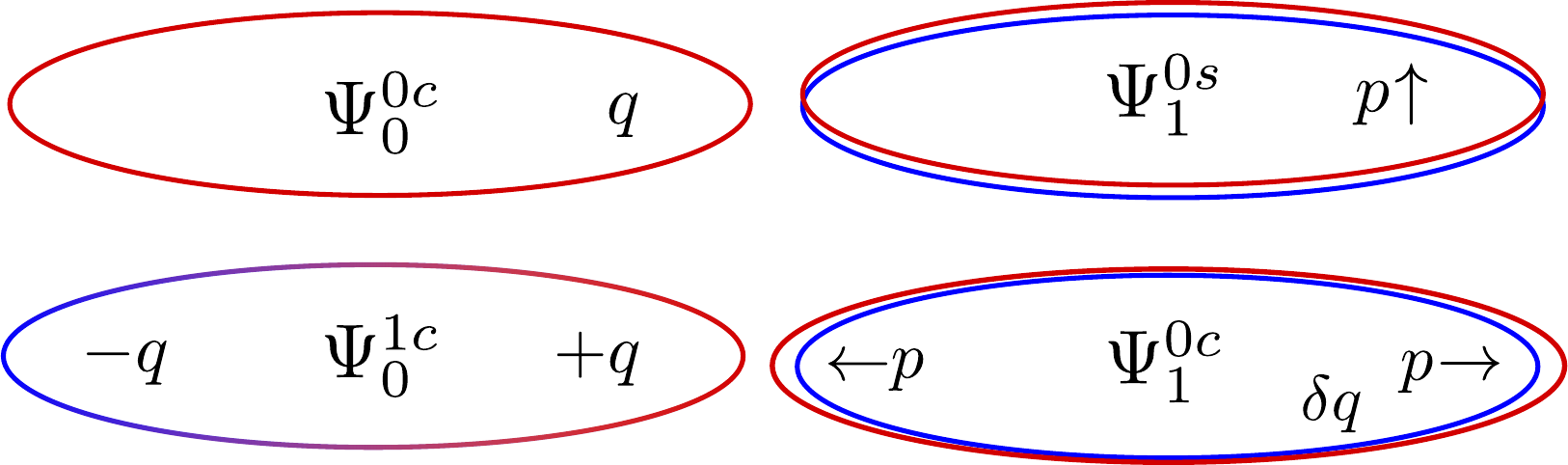}
	\caption{Nature of the singularities of ring-toroidal harmonics in terms of positive and negative rings of charge. $q$ represents a positive charge, $\delta q$ an infinitesimal charge, and $p$ a dipole moment.} \label{rings}
\end{figure}

\begin{figure}[h]
	\includegraphics[scale=1.5]{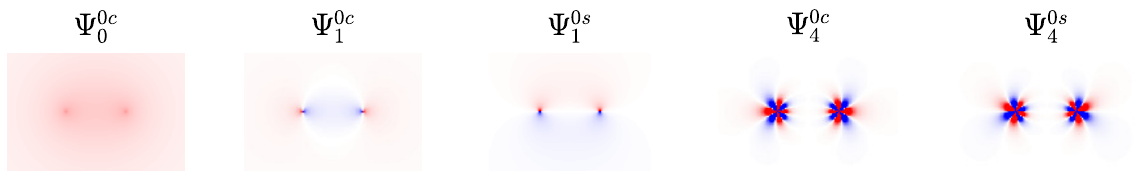}\\
	\includegraphics[scale=1.5]{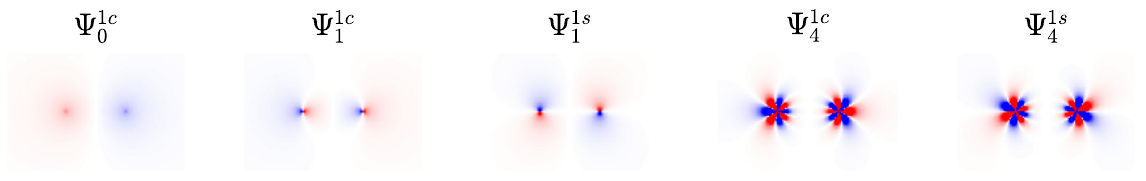}\\
	\includegraphics[scale=1.5]{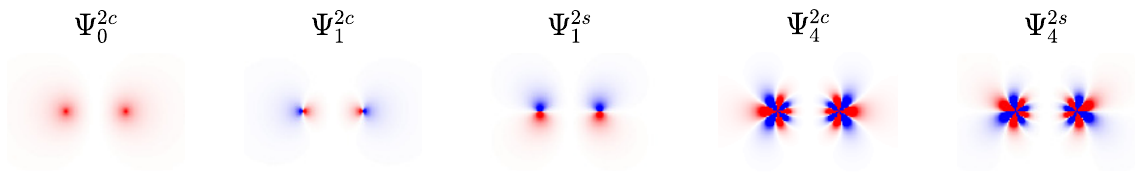}
	\caption{Plots of the standard axial toroidal harmonics for $n=0,1,4$ and $m=0,1,2$. The harmonics are scaled by different linear factors for visualization.} \label{ringplots}
\end{figure}

\begin{figure}[h]
	\includegraphics[scale=.9]{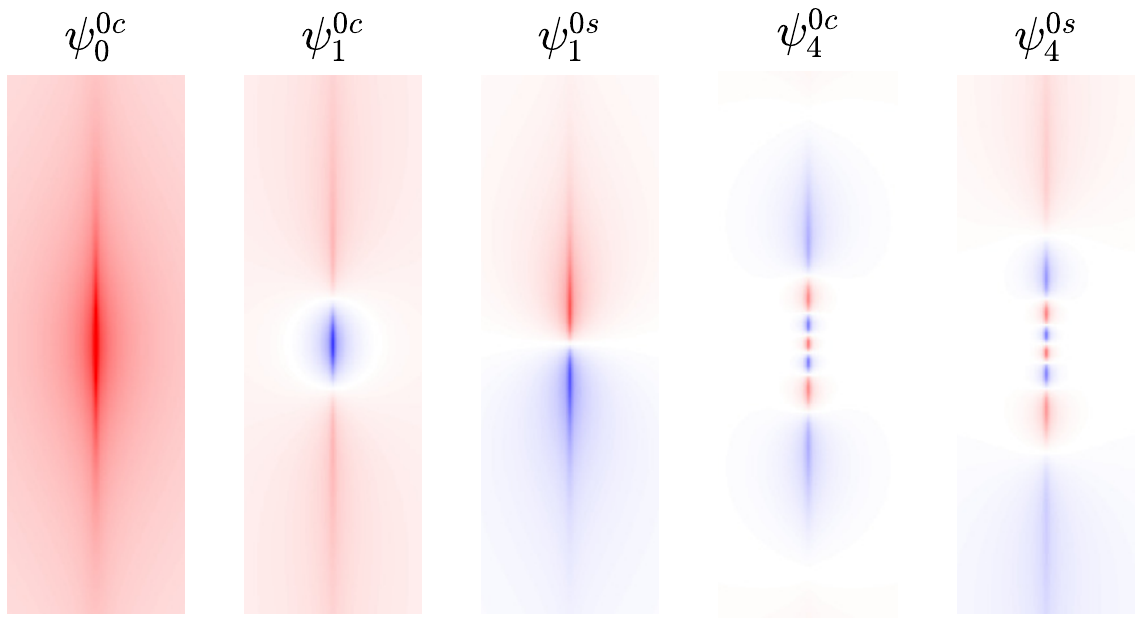}\\
	\includegraphics[scale=.9]{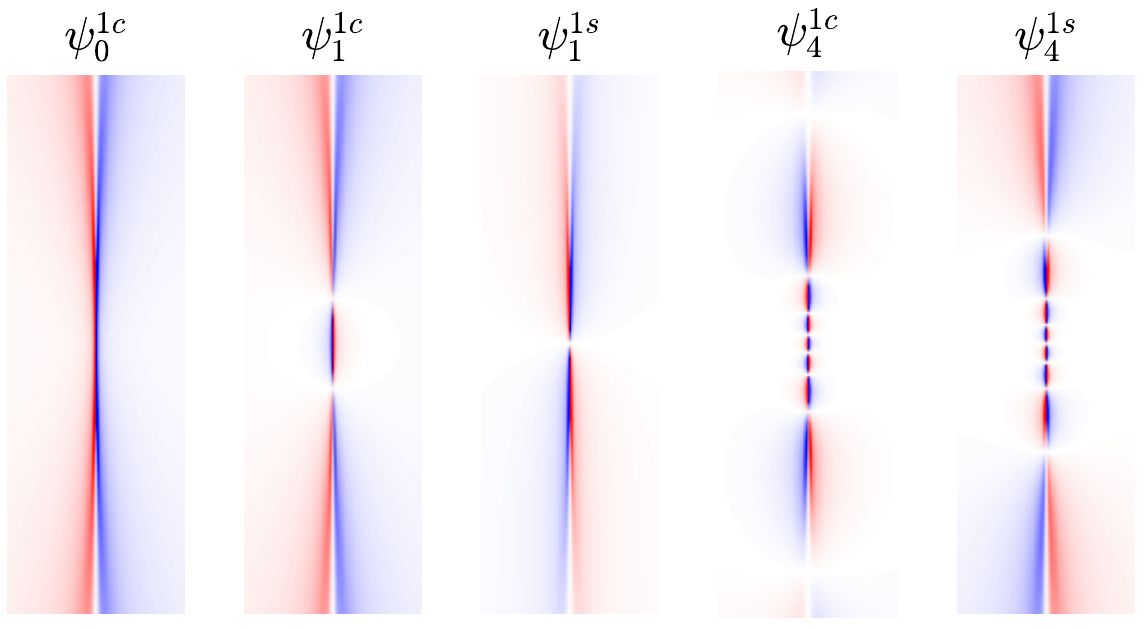}\\
	\includegraphics[scale=.9]{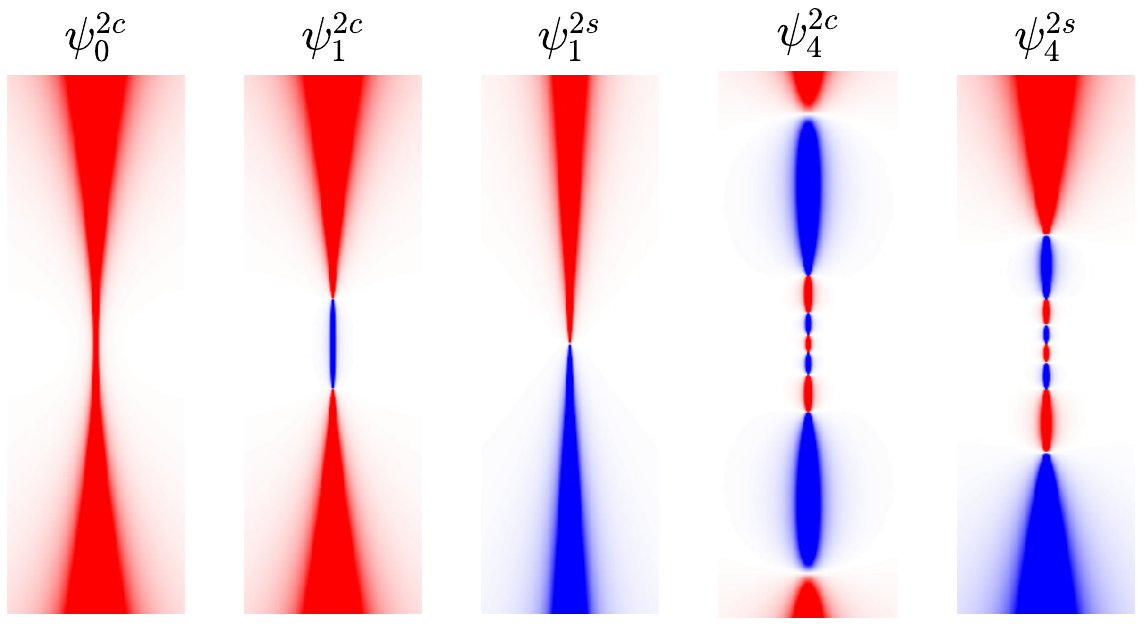}
	\caption{Plots of the standard axial toroidal harmonics for $n=0,1,4$ and $m=0,1,2$. The focal ring radius is 1/4 the width of each plot. The color scale is [-10,10] for $m=0$, [-50,50] for $m=1$ and [-500,500] for $m=2$.} \label{axialplots}
\end{figure}

Laplace's equation is partially separable in toroidal coordinates, meaning that solutions can be written as a product of functions of each coordinate, and a coordinate-dependent prefactor. 
There are two variations of toroidal harmonics that essentially differ by normalization. What we will call the `standard' toroidal harmonics are more commonly used, while the `alternative' toroidal harmonics were first discussed in 2006 \cite{Andrews2006}. 
These are:\\
\begin{alignat}{6}
&\text{`standard ring harmonics': }~ &\Psi_n^{m\substack{c\\s}}&=&\Delta P_{n-1/2}^m(\beta)&{\cos\atop\sin} n\eta e^{\pm im\phi} , &\\[2ex]
&\text{`standard axial harmonics': }~ &\psi_n^{m\substack{c\\s}}&=&\Delta Q_{n-1/2}^m(\beta)&{\cos\atop\sin} n\eta e^{\pm im\phi} ,&\\[2ex]
&\text{`alternate ring harmonics': }~&\Phi_n^{m\substack{c\\s}}&=&\sqrt{\frac{a}{\rho}}Q_{m-1/2}^n(\chi) & {\cos\atop\sin} n\eta e^{\pm im\phi},&\\[2ex]
&\text{`alternate axial harmonics': }~&\phi_n^{m\substack{c\\s}}&=&\sqrt{\frac{a}{\rho}}P_{m-1/2}^n(\chi)  & {\cos\atop\sin} n\eta e^{\pm im\phi},&
\end{alignat}
$$\text{with }~ \Delta=\sqrt{2(\beta-\cos\eta)}.  $$
The superscript $s$ or $c$ refers to whether the sine or cosine of $n\eta$ is taken.
$P_{n-1/2}^m(\beta)$ and $Q_{n-1/2}^m(\beta)$ are Legendre functions of half-integer degree, also called toroidal functions; see appendix for methods of computation. Note the interchange of the indicies of the Legendre functions. 
We will use the term `ring harmonics' for the ones containing $P_{n-1/2}^m(\beta)$ or $Q_{m-1/2}^n(\chi)$  as they are singular on the focal ring, and `axial harmonics' for the ones containing $Q_{n-1/2}^m(\beta)$ or $P_{m-1/2}^n(\chi)$ as they are singular on the entire $z$ axis. Schematics of the source distributions for the ring harmonics are shown in figure \ref{rings}. The ring harmonics are also plotted in figure \ref{ringplots} on the plane $y=0$, which confirms these source distributions for the four cases shown. The singularities for the axial harmonics all lie on the $z$-axis, with $m$ infinitesimally close lines with charge distribution of alternating sign. The nature of the source charge distributions for these functions are apparent from the plots on the plane $y=0$, shown in figure \ref{axialplots}.

In \cite{majic2019toroidalTmat}, expressions were derived for the toroidal harmonics in terms of integrals over their source distribution. For the ring harmonics, the expressions for the low orders $n=0,1$ are:
\begin{align}
\Psi_0^{mc}
&=\frac{(2m-1)!!}{(-2)^m\pi}\int_0^{2\pi}\!\!\frac{e^{im\phi'}a\d\phi'}{\sqrt{r^2+a^2-2\rho a\cos(\phi\!-\!\phi')}}. \label{int Psi n=0}
\end{align}
\begin{align}
\Psi_1^{ms}=&\frac{-a\pd_z}{m-1/2}\Psi_0^{mc}\nonumber\\
=&\frac{(2m-3)!!}{(-2)^m\pi}\int_0^{2\pi}\!\!\frac{-2a^2 ze^{im\phi'}\d\phi'}{(r^2+a^2-2\rho a\cos(\phi\!-\!\phi'))^{3/2}} \label{int Psi n=1 s}
\end{align}
For higher orders, the multiline nature of the ring singularities makes these expressions complicated, so they are probably best expressed as partial derivatives of the ring charges, in the way that a point dipole along $z$ is the derivative in $z$ direction of a point charge. The expressions are
\begin{align}
\Psi_n^{mc}&=\frac{(-)^n}{2}c_n^m(r\pd_r)\Psi_0^{mc} \label{rec toroidal cos}\\
\Psi_n^{ms}&=\frac{(-)^{n+1}}{4}\frac{s_n^m(r\pd_r)}{r\pd_r+1}\Psi_1^{ms} \label{rec toroidal sin}
\end{align}
The function $c_n^m(r\pd_r)$ is equal to $c_{nk}^m$ with $k\rightarrow r\pd_r$ and similarly for $s_n^m$, where $c_{nk}^m$ and $s_{nk}^m$ are polynomials in $k$ defined in \eqref{rec sc}. The multiline rings are apparent in the potential plots in figure \ref{ringplots}.\\

The axial toroidal harmonics are produced by sources on the $z$-axis:
\begin{align}
\psi_n^{mc}=&(2m-1)!! a\bigg(\frac{-a\rho}{2}\bigg)^m e^{im\phi}  \int_{-\infty}^\infty  \frac{(v^2+1)^{m-1/2}~T_n\big(\frac{v^2-1}{v^2+1}\big) }{(\rho^2+(z-av)^2)^{m+1/2}}\d v  \label{line}\\
\psi_n^{ms}=&(2m-1)!! a\bigg(\frac{-a\rho}{2}\bigg)^m e^{im\phi}  \int_{-\infty}^\infty  \frac{(v^2+1)^{m-1/2}~ \frac{2 v}{v^2+1}U_{n-1}\big(\frac{v^2-1}{v^2+1}\big)}{(\rho^2+(z-av)^2)^{m+1/2}}\d v  \label{line}
\end{align}
For $m=0$ the total charge is infinite. For $n=m=0$ the charge decreases as $v\rightarrow\pm\infty$ (away from the origin), but for all other cases, the charge \text{increases} as $v\rightarrow\pm\infty$. This behavior is seen in the plots in figure \ref{axialplots}. 

The null regions of both the ring and axial harmonics lie on the spheres $\eta=k\pi$, $k=1..n$ for the $\sin n\eta$ type, and $\eta=(k+1/2)\pi$, $k=1..n$ for the $\cos n\eta$ type, as is mildly apparent in figures \ref{ringplots} and \ref{axialplots}.

The standard and alternate harmonics are in fact identical up to a prefactor of $n$ and $m$. This can be seen from the Whipple formulae, which relate the Legendre functions $P$ and $Q$ of half integer degree. Expressed in toroidal coordinates the Whipple formulae are
\begin{align}
\Delta P_{n-1/2}^m(\beta)=\frac{(-)^n2/\sqrt{\pi}}{\Gamma(n-m+\frac{1}{2})}\sqrt{\frac{a}{\rho}}Q_{m-1/2}^n(\chi) \label{WhipP}
\\
\Delta Q_{n-1/2}^m(\beta)=\frac{(-)^n\pi\sqrt{\pi}}{\Gamma(n-m+\frac{1}{2})}\sqrt{\frac{a}{\rho}}P_{m-1/2}^n(\chi). 
\label{WhipQ}
\end{align}
Note for half integer arguments
\begin{align}
\Gamma(n+\frac{1}{2})=\sqrt{\pi}\frac{(2n-1)!!}{2^n}, \qquad \Gamma(-n+\frac{1}{2})=\sqrt{\pi}\frac{(-2)^n}{(2n-1)!!} \qquad (n\geq0).
\end{align}

\subsection{Green's function expansions}
We will utilize three different expansions of the inverse distance in deriving the relationships between spherical and toroidal harmonics. For points $\b r_1$ and $\b r_2$ with $r_1<r_2$ , the spherical harmonic expansion of Green's function is
\begin{align}
\frac{1}{|\b r_1-\b r_2|}&=\sum_{m=0}^\infty\epsilon_m(-)^m\sum_{n=m}^\infty\frac{r_1^n}{r_2^{n+1}}P_n^m(u_1)P_n^{-m}(u_2)\cos m(\phi_1-\phi_2), \qquad  \epsilon_m=\begin{cases}1 \quad &m=0 \\ 2 &m>0 \end{cases} \label{GFP}
\end{align}
 Note that $P_n^{-m}=(-)^m\frac{(n-m)!}{(n+m)!}P_n^m$ and here we define $P_k^m(u)$ without the phase factor $(-)^m$.\\
In terms of toroidal harmonics for $\beta_1<\beta_2$ (point 2 closer to the focal ring)\cite{Morse1953} :
\begin{align}
\frac{1}{|\b r_1-\b r_2|}
&=\frac{\Delta_1\Delta_2}{2\pi a}\sum_{n=-\infty}^\infty\sum_{m=-\infty}^\infty P_{n-1/2}^m(\beta_1)Q_{n-1/2}^{-m}(\beta_2)\exp[in(\eta_1-\eta_2)+im(\phi_1-\phi_2)] \nonumber \\
&=\frac{\Delta_1\Delta_2}{2\pi a}\sum_{n=0}^\infty \sum_{m=0}^\infty\epsilon_n\epsilon_m  P_{n-1/2}^m(\beta_1)Q_{n-1/2}^{-m}(\beta_2)\cos n(\eta_1-\eta_2) \cos m(\phi_1-\phi_2). \label{GFQP}
\end{align}

Also we have the 'cylindrical' expansion which converges in all space\cite{Cohl1999}:
\begin{align}
\frac{1}{|\b r_1-\b r_2|}=\frac{1}{\pi\sqrt{\rho_1\rho_2}}\sum_{m=0}^\infty\epsilon_m Q_{m-1/2}(\bar\chi)\cos m(\phi_1-\phi_2), \qquad\qquad \bar\chi=\frac{\rho_1^2+\rho_2^2+(z_1-z_2)^2}{2\rho_1\rho_2} \label{GFQ}
\end{align}
Which converges for all $\b r_1\neq \b r_2$.

\section{Relationships between spherical and toroidal harmonics}
\subsection{Expansions of toroidal harmonics}
We first relate the spherical harmonics to the alternate toroidal harmonics, then use the Whipple formulae to transform the identities to relate to the standard toroidal harmonics. The $e^{i m\phi}$ factors are omitted as they can be tagged on at the end. 

We can find the expansion of ring harmonics with $n=0$ in terms of spherical harmonics by equating the different expansions of Green's function.
Evaluating the spherical and cylindrical expansions of Green's function, \eqref{GFP} and \eqref{GFQ}, both at $\rho_2=a,z_2=0$ ($\Rightarrow u_2=0$, $r_2=a$, $\bar\chi=\chi$), and equating each $m^{th}$ term in the sum we have
\begin{align}
\sqrt{\frac{a}{\rho}}
Q_{m-1/2}(\chi)=\pi\sum_{k=m}^\infty (-)^mP_k^{-m}(0)\left(\frac{r}{a}\right)^kP_k^m(u) \qquad r<a\\
\text{with }P_k^{-m}(0)=\begin{dcases}
(-)^{(k+m)/2}\frac{(k-m-1)!!}{(k+m)!!} \quad &k+m\text{ even}\\ 0 &k+m\text{ odd.}
\end{dcases} \label{relation1}
\end{align}

Or the expansion for $r>a$ can be found by setting $\rho_1=a, z_1=0$:
\begin{align}
\sqrt{\frac{a}{\rho}}
Q_{m-1/2}(\chi)=\pi\sum_{k=m}^\infty (-)^mP_k^{-m}(0)\left(\frac{a}{r}\right)^{k+1}P_k^m(u). \qquad r>a
\end{align}
These series only contain terms with $k+m$ and are therefore symmetric about $z$. Note that $\sqrt{\frac{a}{\rho}}Q_{m-1/2}(\chi)\cos(m\phi)$ is the potential of a thin ring with sinusoidal charge distribution.
\footnote{Note that there are two solutions for the $\eta$ dependence as it is a second order differential equation, but the toroidal functions containing $\sin \eta)$ for $n=0$ are zero because $\sin0\eta=0$. Actually the second independent solution for $n=0$ has $\eta$ dependence of just $\eta$, but this cannot be used as it has a discontinuity in space at $\eta=0,\pi$.}

The result \eqref{relation1} has been generalized to the Helmholtz equation (harmonic time dependence), as a spherical wave function expansion of a circular ring with current distribution expressed as a Fourier series \cite{Hamed2014}.

The $n=1$ toroidal harmonics can be obtained from these $n=0$ expansions by applying the differential operators $\partial_z$ and $r\partial_r$, which both preserve harmonicity. The $n=0$ functions are the potential of a thin ring of charge with an azimuthal charge density of $e^{im\phi}$, but with no variation in the $z$ or $\rho$ directions. The $n=1$ functions are the potential of a ring of dipoles pointing in the $z$ or $\rho$ directions. Differentiating a point charge in the $z$ direction produces a dipole along $z$, so applying $\pd_z$ to a ring will produce a ring of dipoles. We use the following derivatives:
\begin{align}
a\frac{\pd\chi}{\pd z}=\sqrt{\chi^2-1}\sin\eta; \quad\qquad
\frac{\d Q_{m-1/2}^n(\chi)}{\d \chi}=\frac{Q_{m-1/2}^{n+1}(\chi)}{\sqrt{\chi^2-1}}+\frac{n\chi}{\chi^2-1}Q_{m-1/2}^n(\chi),
\end{align}
and that $\pd_z$ is also a ladder operator for the spherical harmonics:
\begin{align}
a\frac{\pd}{\pd z}\left[\left(\frac{r}{a}\right)^nP_n^m(u)\right]&=(n+m)\left(\frac{r}{a}\right)^{n-1}P_{n-1}^m(u) ,\\
a\frac{\pd}{\pd z}\left[\left(\frac{a}{r}\right)^{n+1}P_n^m(u)\right]&=-(n-m+1)\left(\frac{a}{r}\right)^{n+2}P_{n+1}^m(u) .
\end{align}
Then applying $a\pd_z$ to the toroidal harmonic expansion for $r<a$ \eqref{relation1}:
\begin{align}
a\frac{\pd}{\pd z}\sqrt{\frac{a}{\rho}}Q_{m-1/2}(\chi)
=\sqrt{\frac{a}{\rho}}Q_{m-1/2}^1(\chi)\sin\eta 
&=\pi\sum_{k=m}^\infty(-)^mP_k^{-m}(0)(k+m)\left(\frac{r}{a}\right)^{k-1}P_{k-1}^m(u) \\
&=\pi\sum_{k=m}^\infty(-)^mP_{k+1}^{-m}(0)(k+m+1)\left(\frac{r}{a}\right)^nP_k^m(u),
\end{align}
and for $r>a$:
\begin{align}
\sqrt{\frac{a}{\rho}}Q_{m-1/2}^1(\chi)\sin\eta=\pi\sum_{k=m}^\infty(-)^mP_{k+1}^{-m}(0)(k+m+1)\left(\frac{a}{r}\right)^{k+1}P_k^m(u).
\end{align}
The series terms are only nonzero for $k+m$ odd, and therefore antisymmetric about $z$.

To produce toroidal harmonics with dipoles oriented outwards from the origin, we can apply $r\pd_r$ ($\partial_\rho$ cannot be used as it doesn't preserve harmonicity). This operator should turn a ring of charge on the $xy$ plane into a ring of dipoles pointing inward. We use the following derivatives:
\begin{align}
r\frac{\pd\chi}{\pd r}=\sqrt{\chi^2-1}\cos\eta; \qquad r\frac{\pd}{\pd r}\sqrt{\frac{a}{\rho}}=\frac{-1}{2}\sqrt{\frac{a}{\rho}}.
\end{align}
Applying $r\pd_r$ to the $n=0$ toroidal harmonic expansion for $r<a$ \eqref{relation1} and rearranging:
\begin{align}
r\frac{\pd}{\pd r}&\sqrt{\frac{a}{\rho}}Q_{m-1/2}(\chi)=\sqrt{\frac{a}{\rho}}\left[Q_{m-1/2}^1(\chi)\cos\eta - \frac{1}{2}
Q_{m-1/2}(\chi)\right]
=\sum_{k=m}^\infty (-)^mP_k^{-m}(0)k\left(\frac{r}{a}\right)^kP_k^m(u) \\
\Rightarrow &\sqrt{\frac{a}{\rho}}Q_{m-1/2}^1(\chi)\cos\eta 
=\sum_{k=m}^\infty (-)^mP_k^{-m}(0)\left(k+\frac{1}{2}\right)\left(\frac{r}{a}\right)^kP_k^m(u).
\end{align}
And for $r>a$:
\begin{align}
\sqrt{\frac{a}{\rho}}Q_{m-1/2}^1(\chi)\cos\eta 
=-\sum_{k=m}^\infty (-)^mP_k^{-m}(0)\left(k+\frac{1}{2}\right)\left(\frac{a}{r}\right)^{k+1}P_k^m(u).
\end{align}
It is interesting that for $m=0$, this series starts from $k=0$ which means that this toroidal harmonic has a monopole moment - its corresponding charge distribution has a net charge.\\

Now we derive the formulae for general $n$ by induction, applying $r\pd_r$ to the $n^{th}$ harmonic (with $\sin n\eta$ or $\cos n\eta$)to derive the expansion for $n+1$. \\
The following formulae are useful:
\begin{align}
r\frac{\pd}{\pd r}\cos n\eta  &= \frac{n\chi}{\sqrt{\chi^2-1}}\sin\eta\sin n\eta , \qquad
r\frac{\pd}{\pd r}\sin n\eta  = \frac{-n\chi}{\sqrt{\chi^2-1}}\sin\eta\cos n\eta ,  \\
2\cos\eta\cos n\eta  &= \cos(n-1)\eta-\cos(n+1)\eta, \qquad 
\cos\eta\cos n\eta +\sin\eta\sin n\eta  = \cos(n-1)\eta \nonumber\\
2\cos\eta\sin n\eta  &= \sin(n-1)\eta+\sin(n+1)\eta, \qquad 
\cos\eta\sin n\eta -\sin\eta\cos n\eta  = \sin(n-1)\eta \\
Q_{m-1/2}^{n+1}(\chi)&=\frac{-2n\chi}{\sqrt{\chi^2-1}}Q_{m-1/2}^n(\chi)+\bigg(m^2-\Big(n-\frac{1}{2}\Big)^2\bigg)Q_{m-1/2}^{n-1}(\chi)
\end{align}
Applying $r\pd_r $ to the the symmetric and antisymmetric harmonics and rearranging to express the result as a sum of harmonic functions we obtain:
\begin{align}
r\frac{\pd}{\pd r}&\sqrt{\frac{a}{\rho}}Q_{m-1/2}^n(\chi){\cos\atop\sin}(n\eta) \nonumber\\
&=\frac{1}{2}\sqrt{\frac{a}{\rho}}\left[Q_{m-1/2}^{n+1}(\chi){\cos\atop\sin}(n+1)\eta-Q_{m-1/2}^n(\chi){\cos\atop\sin}n\eta +\left(m^2-\Big(n-\frac{1}{2}\Big)^2\right)Q_{m-1/2}^{n-1}(\chi){\cos\atop\sin}(n-1)\eta \right] \label{rdrQ}
\end{align}
Now assume the symmetric and antisymmetric toroidal harmonics can be expanded as regular spherical harmonics for $r<a$, as
\begin{align}
\sqrt{\frac{a}{\rho}}Q_{m-1/2}^n(\chi)\cos n\eta &= \pi(-)^m\sum_{k=m}^\infty C_{nk}^mP_k^{-m}(0)\left(\frac{r}{a}\right)^kP_k^m(u), \qquad r<a\\
\sqrt{\frac{a}{\rho}}Q_{m-1/2}^n(\chi)\sin n\eta &= \pi(-)^m\sum_{k=m}^\infty S_{nk}^mP_{k+1}^{-m}(0)\left(\frac{r}{a}\right)^kP_k^m(u). \qquad r<a
\end{align}
Plugging these expansions in to \eqref{rdrQ} and rearranging gives a recurrence relation for $C_{nk}^m$ and $S_{nk}^m$:
\begin{align}
C_{n+1,k}^m&=(2k+1)C_{nk}^m + \left(\Big(n-\frac{1}{2}\Big)^2-m^2\right)C_{n-1,k}^m, \\
S_{n+1,k}^m&=(2k+1)S_{nk}^m + \left(\Big(n-\frac{1}{2}\Big)^2-m^2\right)S_{n-1,k}^m, \label{rec CS}
\end{align}
with initial values that we have determined above for the $n=0,1$ expansions:
\begin{align}
C_{0,k}^m=1, \qquad C_{1,k}^m=k+\frac{1}{2}, \qquad S_{0,k}^m=0,\qquad S_{1,k}^m=k+m+1.
\end{align}
The first few orders for $m=0$ are:
\begin{align*}
C_{2,k}^0&=2(k^2+k+\frac{3}{8}) \\ 
C_{3,k}^0&=2(2k+1)(k^2+k+\frac{15}{16}) \\ 
C_{4,k}^0&=2(4k^4+8k^3+15k^2+11k+\frac{105}{32}) \\
C_{5,k}^0&=2(2k+1)(4k^4+8k^3+\frac{109}{4}k^2+\frac{93}{4}k+\frac{945}{64}) \\
\vspace{.5cm} \\
S_{2,k}^0&=(k+1)(2k+1) \\
S_{3,k}^0&=4(k+1)(k^2+k+\frac{13}{16}) \\
S_{4,k}^0&=4(k+1)(2k+1)(k^2+k+\frac{76}{32}) \\
S_{5,k}^0&=4(k+1)(4k^4+8k^3+\frac{107}{4}k^2+\frac{91}{4} k+\frac{789}{64})
\end{align*}
$C_{nk}^m$ and $S_{nk}^m$ can be computed accurately via the forward recurrence \eqref{rec CS}.\\

For $r>a$ the expansions look almost identical:
\begin{align}
\sqrt{\frac{a}{\rho}}Q_{m-1/2}^n(\chi)\cos n\eta &= \pi(-)^{n+m} \sum_{k=m}^\infty C_{nk}^mP_k^{-m}(0)\left(\frac{a}{r}\right)^{k+1}P_k^m(u) \qquad r>a\\
\sqrt{\frac{a}{\rho}}Q_{m-1/2}^n(\chi)\sin n\eta &= \pi(-)^{n+m+1}\sum_{k=m}^\infty S_{nk}^mP_{k+1}^{-m}(0)\left(\frac{a}{r}\right)^{k+1}P_k^m(u) \qquad r>a
\end{align}

Finally we apply the Whipple formulae \eqref{WhipQ} to express these expansions in terms of standard toroidal harmonics:
\begin{align}
\Delta P_{n-1/2}^m(\beta)\cos n\eta &= 2(-)^m \sum_{k=m}^\infty c_{nk}^mP_k^{-m}(0)
\begin{dcases}
(-)^n\left(\frac{r}{a}\right)^kP_k^m(u) \qquad &r<a\\
\left(\frac{a}{r}\right)^{k+1}P_k^m(u) \qquad &r>a\\
\end{dcases}\\
\Delta P_{n-1/2}^m(\beta)\sin n\eta &= 2(-)^m\sum_{k=m}^\infty s_{nk}^mP_{k+1}^{-m}(0)
\begin{dcases}
(-)^n\left(\frac{r}{a}\right)^kP_k^m(u) \qquad &r<a\\
-\left(\frac{a}{r}\right)^{k+1}P_k^m(u) \qquad &r>a
\end{dcases} \label{toroidal_spherical}
\end{align}
where
\begin{align}
c_{nk}^m=\frac{\sqrt{\pi}}{\Gamma(n-m+\frac{1}{2})}C_{nk}^m, \qquad
s_{nk}^m=\frac{\sqrt{\pi}}{\Gamma(n-m+\frac{1}{2})}S_{nk}^m,
\end{align}
which both follow the same recurrence
\begin{align}
&\bigg(n-m+\frac{1}{2}\bigg) c_{n+1,k}^m = (2k+1)c_{nk}^m + \bigg(n+m-\frac{1}{2}\bigg)c_{n-1,k}^m,  \label{rec sc}
\end{align}
with initial conditions
\begin{align}	
&c_{0,k}^m=\frac{\sqrt{\pi}}{\Gamma(-m+\frac{1}{2})}
,\quad c_{1,k}^m=\left(k+\frac{1}{2}\right)\frac{\sqrt{\pi}}{\Gamma(-m+\frac{3}{2})} ,
\qquad s_{0,k}^m=0, \quad s_{1,k}^m=(k+m+1)\frac{\sqrt{\pi}}{\Gamma(-m+\frac{3}{2})}.
\end{align}

In Ref. \cite{erofeenko1983addition}, these coefficients were expressed together using complex notation. There the operator $\pd_z$ was used to derive a triangular recurrence over $n$ and $k$:
\begin{align}
2i(k-m)h_{n,k-1}^m = \Big(n-m+\frac{1}{2}\Big)h_{n+1,k}^m - 2nh_{nk}^m +\Big(n+m-\frac{1}{2}\Big)h_{n-1,k}^m,
\end{align}
where $h_{nk}^m=(-)^k[c_{nk}^mP_k^{-m}(0) + i s_{nk}^mP_{k+1}^{-m}(0) ]$.\\
For more properties and explicit forms of these coefficients, see \cite{majic2019toroidalTmat}.

\subsection{Expansions of spherical harmonics}
The expansions of spherical harmonics in terms of axial toroidal harmonics can be found from inserting the spherical expansion of the ring harmonics in to the toroidal harmonic expansion of Green's function \eqref{GFQP}, and then equating this to the spherical expansion. We first use a trigonometric identity to express the toroidal Green's function expansion as 
\begin{align}
\frac{1}{|\b r_1-\b r_2|}
&=\frac{\Delta_1\Delta_2}{2\pi a}\sum_{m=0}^\infty \sum_{k=0}^\infty\epsilon_k\epsilon_m  P_{k-1/2}^m(\beta_1)Q_{k-1/2}^{-m}(\beta_2)[\cos(k\eta_1)\cos(k\eta_2)+\sin(k\eta_1)\sin(k\eta_2)]\cos m(\phi_1-\phi_2) .
\end{align}
We can then insert the regular expansions for point 2 \eqref{toroidal_spherical}, and equate this to the spherical expansion of Green's function \eqref{GFP}:
\begin{align}
\frac{1}{|\b r_1-\b r_2|}&=\frac{\Delta_2}{a}\sum_{m=0}^\infty \sum_{k=0}^\infty\epsilon_k\epsilon_m  Q_{k-1/2}^{-m}(\beta_2)\frac{(-)^k/\sqrt{\pi}}{\Gamma(k-m+\frac{1}{2})}\sum_{n=m}^\infty \left[\cos(k\eta_2) C_{kn}^mP_n^{-m}(0)+ \sin(k\eta_2)S_{kn}^mP_{n+1}^{-m}(0)\right]
\nonumber\\
&\hspace{8cm}\times(-)^m\left(\frac{r_1}{a}\right)^nP_n^m(u_1)\cos m(\phi_1-\phi_2) \\ 
&=\sum_{m=0}^\infty\sum_{n=m}^\infty\epsilon_m(-)^m\frac{r_1^n}{r_2^{n+1}}P_n^{-m}(u_1)P_n^m(u_2)\cos m(\phi_1-\phi_2).
\end{align}
Equating the coefficients of the spherical harmonics of point 1 for each degree and order, and noting that $Q_{n-1/2}^{-m}=(-)^m\frac{\Gamma(n-m+\frac{1}{2})}{\Gamma(n+m+\frac{1}{2})}Q_{n-1/2}^m$, we find the expansion of irregular spherical harmonics ($P_n^m(0)$ and $P_{n+1}^m(0)$ are alternately zero for $n+m$ even or odd):
\begin{align}
\left(\frac{a}{r}\right)^{n+1}P_n^{m}(u)=\frac{\Delta}{\pi}(-)^m\begin{dcases}
P_n^m(0)\sum_{k=0}^\infty \epsilon_k(-)^k c_{kn}^{-m}Q_{k-1/2}^m(\beta)\cos k\eta \qquad &n+m\text{ even} \\
2P_{n+1}^m(0)\sum_{k=1}^\infty (-)^k s_{kn}^{-m}Q_{k-1/2}^m(\beta)\sin k\eta \qquad &n+m\text{ odd.}
\end{dcases} \label{irr_spherical_toroidal}
\end{align}
For the regular spherical harmonics, a similar derivation gives
\begin{align}
\left(\frac{r}{a}\right)^nP_n^{m}(u)=\frac{\Delta}{\pi}(-)^m\begin{dcases}P_n^m(0)
\sum_{k=0}^\infty\epsilon_k c_{kn}^{-m}Q_{k-1/2}^m(\beta)\cos k\eta \qquad &n+m\text{ even} \\
-2P_{n+1}^m(0)\sum_{k=1}^\infty s_{kn}^{-m}Q_{k-1/2}^m(\beta)\sin k\eta \qquad &n+m\text{ odd,}
\end{dcases} \label{reg_spherical_toroidal}
\end{align}
where
\begin{align}
c_{kn}^{-m}
=\frac{\Gamma(k-m+1/2)}{\Gamma(k+m+1/2)}c_{kn}^{m}, \qquad  
 s_{kn}^{-m}
 =\frac{\Gamma(k-m+1/2)}{\Gamma(k+m+1/2)}\frac{k-m+1}{k+m+1}s_{kn}^{m},
\end{align}

Again these can be re-expressed in terms of alternative toroidal harmonics using the Whipple formulae.
For $n=0$, \eqref{reg_spherical_toroidal} is Heine's expansion of a constant, and for $n=1$, $m=0,1$, has been used in the context of low frequency plane wave scattering \cite{venkov2007low}. The rest are presumably unknown.
Again the coefficients $c_{kn}^m, s_{kn}^m$ do not have simple closed forms (for fixed $n$ and $m$), but some low orders are
\begin{align}
c_{k0}^0=1, \qquad 
s_{k1}^0=4k,  \qquad
c_{k2}^0=4k^2+1, \qquad
s_{k3}^0=\frac{8}{9}(4k^3+5k), \qquad 
c_{k1}^1=\frac{1}{2}(4k^2-1), \qquad
c_{k1}^{-1}=2. 
\end{align}
These can be proven by substitution into the recurrence \eqref{rec sc}.
$c_{kn}^m$ only appear in the series for $n+m$ even, while $s_{kn}^m$ only for $n+m$ odd. \\

\subsection{Region of convergence}
We can determine the boundary of convergence of expansions \eqref{reg_spherical_toroidal} and \eqref{irr_spherical_toroidal} from the behaviour of the $k^{th}$ term in the series as $k\rightarrow\infty$. The Legendre functions grow as \cite{lebedev1965special} pg 191 (\eqref{limP} is presented for later):
\begin{align}
\lim_{k\rightarrow\infty} P_{k-1/2} (\cosh\xi) =& \frac{e^{k\xi}}{\sqrt{(2k-1)\sinh\xi}} \label{limP}\\
\lim_{k\rightarrow\infty} Q_{k-1/2} (\cosh\xi) =& \frac{\sqrt{\pi}e^{-k\xi}}{\sqrt{(2k-1)\sinh\xi}}  \label{limQ}
\end{align}
The series coefficients $c_{kn}^m$ and $s_{kn}^m$ are bounded by the sequence $e_{k+1}=(2n+1)/k e_k + e_{k-1}$ (with the same initial values), which itself grows slower than $k^{2n+1}$. $\Gamma(k+m-1/2)/k!$ is also bounded by a polynomial in $k$ (degree $m$). All together the series decreases exponentially 
and converges everywhere except $\xi=0$ (the $z$-axis and at $r=\infty$).

Numerically, the expansions \eqref{irr_spherical_toroidal} and \eqref{reg_spherical_toroidal} converge slowly away from the focal ring - near the $z$ axis and far from the origin. In these cases the series terms grow significantly in magnitude before converging, which sacrifices accuracy due to catastrophic cancellation between the terms. \\

\subsection{Existence of expansions}
\begin{figure}[h]
	\includegraphics[scale=.57]{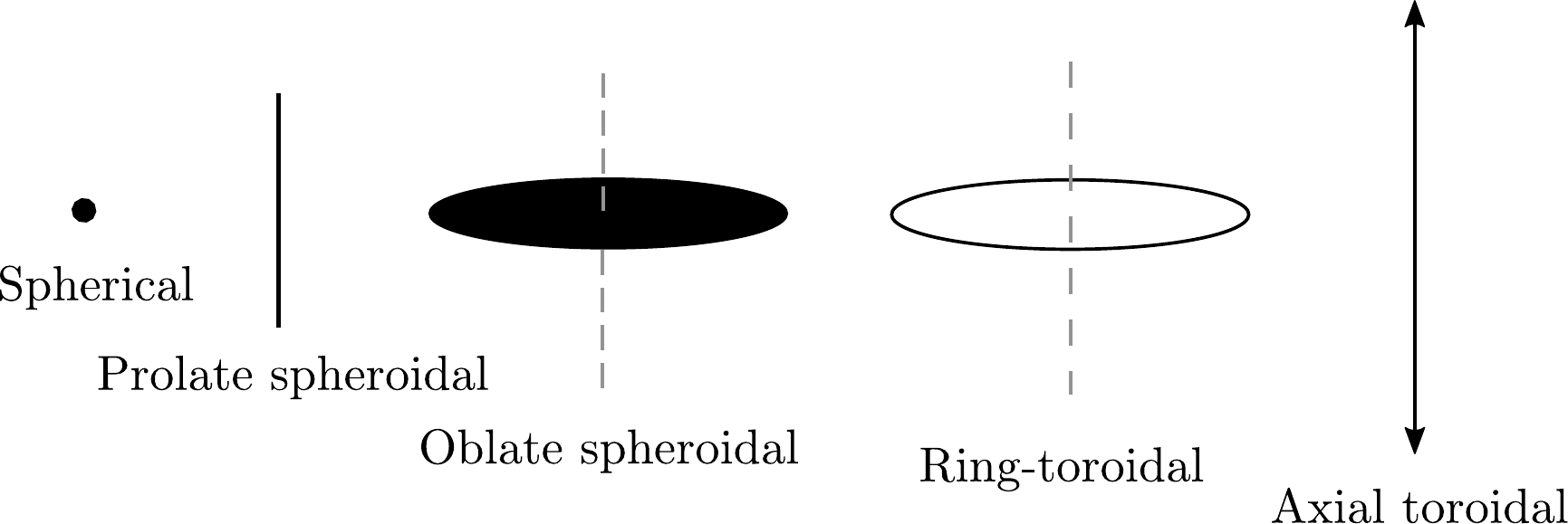}
	\caption{Singular regions of the external spherical, external prolate and oblate spheroidal, ring-toroidal, and axial toroidal harmonics. The dashed lines represent the $z$-axis.} \label{singularities}
\end{figure}
We can compare these expansions to those between spherical and spheroidal harmonics, which are much more similar in terms of their singularities, shown in figure \ref{singularities} - point singularity for the external spherical harmonics, finite line singularity for the external prolate spheroidal harmonics, and disc singularity for the oblate spheroidal harmonics. The external spheroidal harmonics decay as $r\rightarrow\infty$, so are expressible as series of external spherical harmonics, and vice versa \cite{Jansen2000}.
Similarly, both prolate and oblate \textit{internal} spheroidal harmonics are finite at the origin and diverge at $r=\infty$, just as do the spherical harmonics, and can be written as a finite sum of internal spherical harmonics, and vice versa. Naturally the internal spheroidal harmonics cannot be expressed in terms of external spherical harmonics or vice versa. 

But the toroidal harmonics do not follow the same notion of internal and external. We have shown that the ring toroidal harmonics can be written as a series of either internal or external spherical harmonics. This is due to the fact that they are finite at both the origin and at $r=\infty$. However the axial toroidal functions are singular at the origin and at infinity, which means that they cannot be expanded as a series of spherical harmonics at all.

Also, neither internal and external spherical harmonics can be expressed as a series of ring harmonics. An intuition for this is that any series of ring harmonics will only converge outside some toroidal boundary, and this boundary must enclose the singularity(s) of the function being expanded. The external spherical harmonics are singular at the origin, so this toroidal boundary must cover the origin, however, this torus will then extend to all space. The internal spherical harmonics cannot be expanded for a similar reason - the torus must extend to all space to cover the `singularity' at $r=\infty$. 

However, both internal and external spherical harmonics can be expanded with toroidal line harmonics, because any series of axial toroidal harmonics will converge \textit{inside} some torus. For the internal spherical harmonics, this toroidal boundary may extend up to infinity since that is where the singularity at $r=0$ or $r\rightarrow\infty$ lies. For external spherical harmonics the toroidal boundary may extend to the origin. And as shown mathematically in the previous section, the toroidal boundary actually does extend to all space. 

To summarize, the expansions relating toroidal and spherical harmonics are all non-invertible.

\section{Charged conducting torus}

We demonstrate the use of these expansions in expressing the potential of a charged conducting torus on a spherical basis.\\
Consider a perfectly conducting torus held at potential $V_0$, with major radius $R_0$ and minor radius $r_0$. 
The focal ring radius and surface parameter $\beta=\beta_0$ are obtained from
\begin{align}
a=
\sqrt{R_0^2-r_0^2}, \qquad \beta_0=
\frac{R_0}{r_0}.
\end{align}

Using toroidal harmonics the solution is given by \cite{belevitch1983torus}:
\begin{align}
V=V_0\frac{\Delta}{\pi}\sum_{n=0}^\infty\epsilon_n\frac{Q_{n-1/2}(\beta_0)}{P_{n-1/2}(\beta_0)}P_{n-1/2}(\beta)\cos n\eta .\label{Vconducting toroidal}
\end{align}
Substituting the expansion of toroidal in terms of spherical harmonics and rearranging the summation order:
\begin{align}
V=\frac{2V_0}{\pi}\sum_{k=0}^\infty\sum_{n=0}^\infty \epsilon_n\frac{Q_{n-1/2}(\beta_0)}{P_{n-1/2}(\beta_0)}c_{nk} P_k(0)
\begin{dcases}
(-)^n\left(\frac{r}{a}\right)^kP_k(u) \qquad &r\rightarrow0 \\
\left(\frac{a}{r}\right)^{k+1}P_k(u)  \qquad &r\rightarrow\infty.
\end{dcases} \label{sph cond}
\end{align}
\begin{figure}[h!]
	\includegraphics[scale=.99]{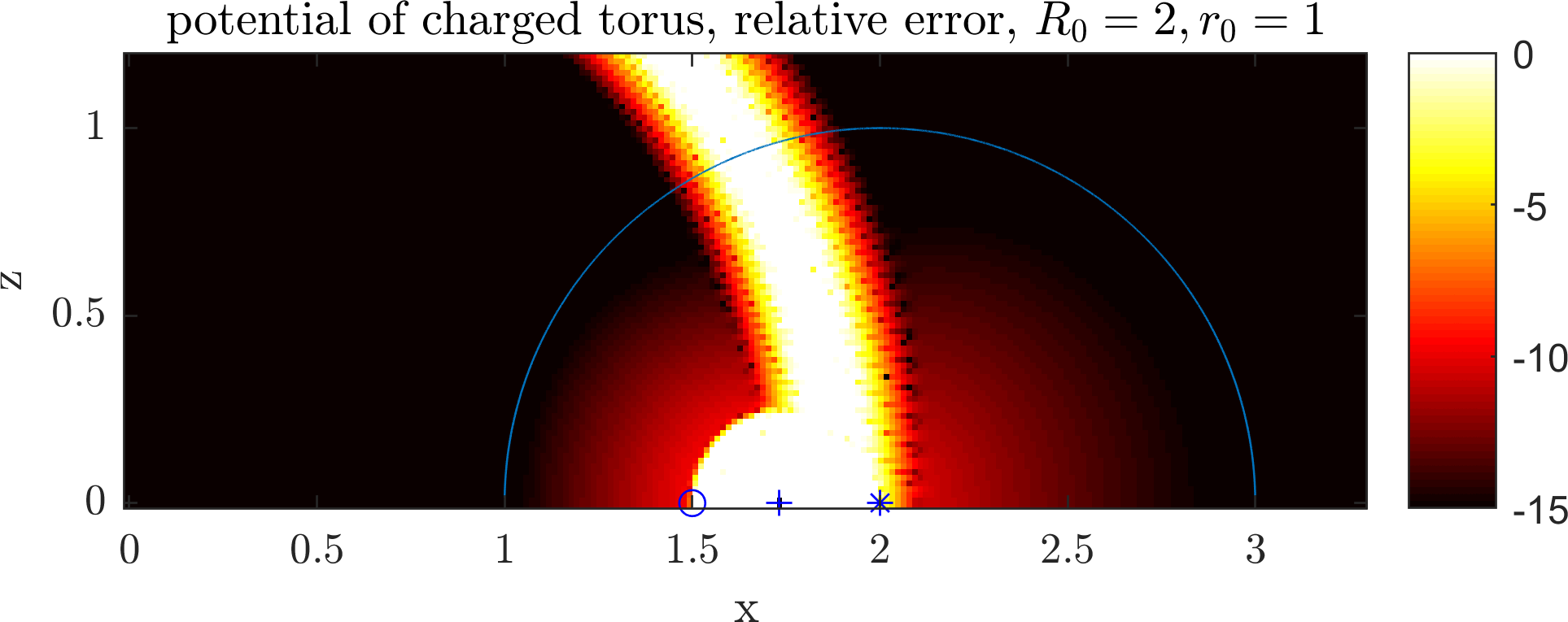}\\~\\
	\includegraphics[scale=.99]{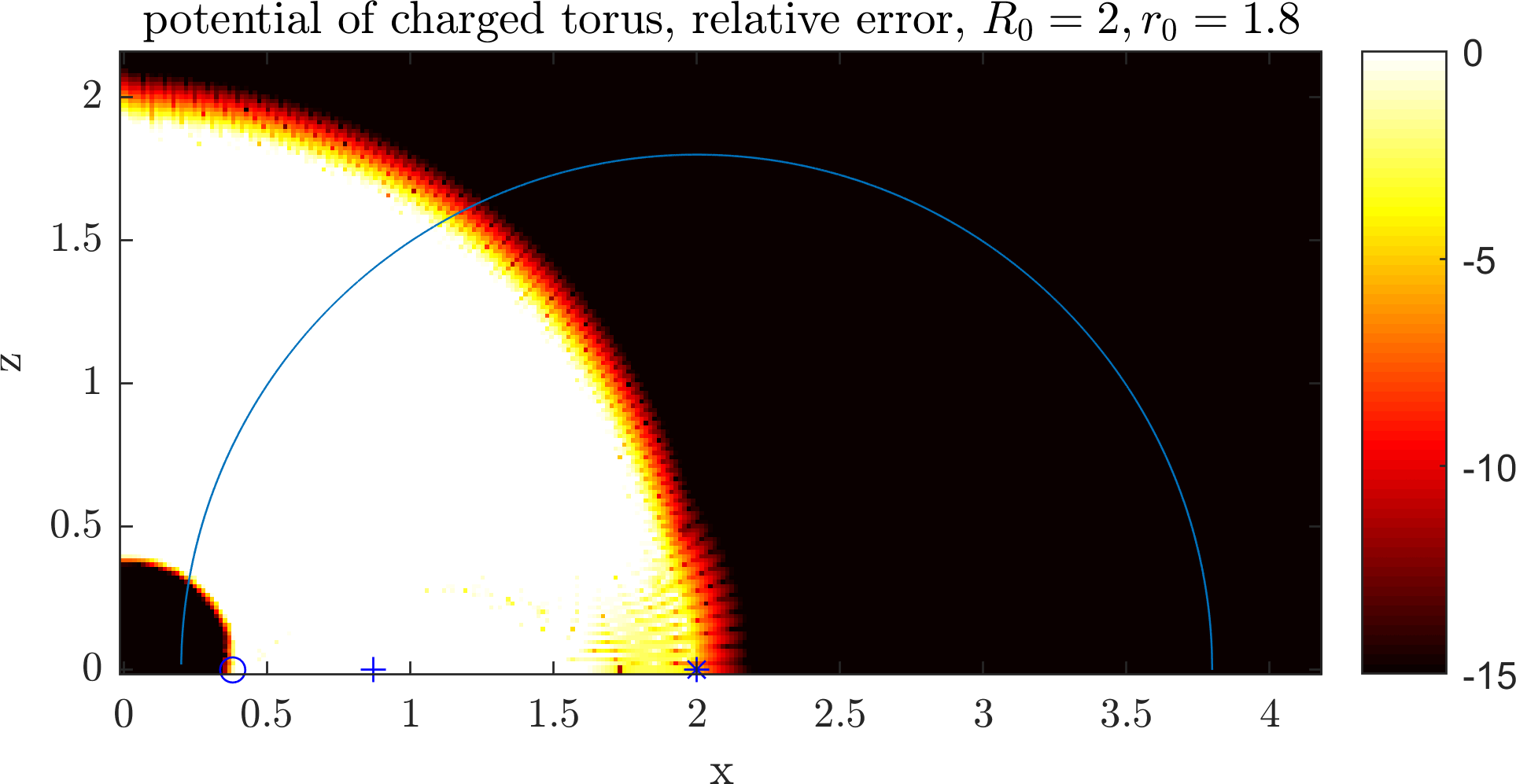}
	\caption{relative error of the potential of a charged conducting torus as computed by the toroidal and spherical series, \eqref{Vconducting toroidal}, \eqref{sph cond}.  The blue circle is the edge of the torus and the points are \color{blue}$\circ$\color{black}: $\rho=R_0-r_0^2/R_0$,  \color{blue}+\color{black}: $\rho=a$,  \color{blue}$\ast$\color{black}: $\rho=R_0$. The toroidal series is computed up to $n=120$ and the spherical up to $k=170$ with $n=120$. The black regions are where both the spherical and toroidal harmonic solutions have converged to floating point accuracy. Increasing the number of terms would shrink the red and yellow regions, but numerical problems start to arise. We see the overlap of the regions of divergence in white - the inner torus and the spherical annulus. } \label{rel err}
\end{figure}

We now briefly study the convergence of these series. It seems difficult to analyze domain of convergence of the spherical series directly from the series coefficients in \eqref{sph cond}, since the coefficients are a series themselves. Perhaps a simpler form of the series coefficients can be obtained via direct integration over the surface, for which boundaries of convergence can be obtained, similar to the approach in \cite{Kondratyev2009,Kondratyev2010} for the uniform solid torus. Anyway, numerical tests in figure \ref{rel err} indicate that the boundary of convergence of the $r\rightarrow\infty$ spherical solution is $r=R_0$, while the boundary of convergence of the $r\rightarrow0$ spherical solution lies somewhere between $r=R_0-r_0^2/R_0$ and $r=a=\sqrt{R_0^2-r_0^2}$, depending on the torus thickness. The potential can only be singular between these two boundaries.

We can also make deductions from the toroidal solution \eqref{Vconducting toroidal}. Using the asymptotic formulas \eqref{limP} and \eqref{limQ}, we can show that the toroidal series converges for $\xi<2\xi_0$, an inner torus with focal radius $a$, extending on the $xy$-plane to the center of the tube at $R_0$. The intrinsic singularity of the potential then lies within this smaller torus. Assuming that the singularity touches the edge of this inner torus (which seems a natural assumption since the series behaves geometrically in $\xi$), this implies that there is an intermediate spherical annulus where the internal and external spherical harmonic series both diverge \eqref{sph cond}.

This problem of divergence is also encountered in the problem of the gravity of a solid torus \cite{Kondratyev2009,Kondratyev2010}. An attempted solution for the potential in the intermediate region is given in terms of a series of both regular and irregular spherical harmonics \cite{Kondratyev2012}, but this fails to give exact results due to the singularity of the potential being in this region. For this problem we expect the same - that the potential in the intermediate annulus cannot be evaluated using spherical harmonic series.

\appendix
\section{Computation of toroidal functions}
The Legendre functions of half-integer degree have the following series representations \cite{Morse1953,rotenberg1960calculation}:
\begin{align}
P_{n-1/2}^m(x)&=\frac{\sqrt{2\pi}(x^2-1)^{m/2}(x+1)^{-n-m-1/2}}{\Gamma(n-m+\frac{1}{2})(2n-1)!!} \sum_{k=0}^\infty\frac{(2(n+m+k)-1)!!(2(n+k)-1)!!}{k!(m+k)!2^{m+k}}\left(\frac{x-1}{x+1}\right)^k\\
Q_{n-1/2}^m(x)&=\pi(-)^m\frac{(x^2-1)^{m/2}}{(2x)^{n+m+1/2}}\sum_{k=0}^\infty\frac{(4k+2n+2m-1)!!}{(2k)!!(2k+2n)!!}\frac{1}{(2x)^{2k}} \qquad x>1, n\geq0, m\geq0
\end{align}
It is often faster to compute the functions by recurrence, using the series to compute the initial values. Matlab functions for computing the half-integer Legendre functions are attached as supplementary material. Advanced methods of calculating these functions are detailed in \cite{Segura2000}. 

\acknowledgements
This research was funded by a Victoria University of Wellington doctoral scholarship.

\bibliography{../libraryH}

\end{document}